# Design of a Compact Monolithic Catadioptric Lens for CubeSat Hyperspectral Imaging Payloads

Iliya Shofman*, Kenya He
University of Toronto Aerospace Team - Space Systems Division
55 St. George Street, Myhal Centre, Room 618, Toronto, ON Canada M5S 0C9

## ABSTRACT

Monolithic catadioptric lenses have emerged as a novel class of compact long-focal-length lenses. Fabricated from a single fused silica substrate with distinct optical prescriptions across annular zones, these systems fold the optical path within the substrate, achieving long effective focal lengths at a fraction of the track length of refractive equivalents. The monolithic architecture and low coefficient of thermal expansion provide inherent athermal performance and preserve optical alignment under the thermal and vibrational loads of nanosatellite launch and orbital operation.

Despite these advantages, commercial offerings of monolithic catadioptric lenses remain limited, and no comprehensive optical design literature review has been published for this lens class. We address this gap with a systematic survey of monolithic catadioptric design variants, including Schmidt-Cassegrain-derived, annular folded, and gradient-index configurations, with emphasis on their suitability for nanosatellite imaging payloads.

Building on this design space, we also present a novel monolithic catadioptric objective lens optimized for a CubeSat pushbroom hyperspectral Earth imaging payload. Pushbroom hyperspectral imagers require the objective lens to be coupled to a precision slit, and maintaining slit-to-focal-plane alignment under launch shock and thermal cycling is a significant optomechanical challenge. The proposed design places the focal plane coincident with the back surface of the substrate, enabling the slit to be lithographically etched directly onto that surface. This approach eliminates slit-to-objective alignment, yielding an integrated, mechanically robust, and thermally stable objective-slit module for pushbroom imaging. We present the completed optical design of a near-diffraction-limited F/3 f=100mm VSWIR monolithic catadioptric lens, assess imaging performance through simulation, and discuss preliminary fabrication and tolerancing considerations.

**Keywords:** monolithic catadioptric lens, lens design, CubeSat, nanosatellite, hyperspectral imaging, SWIR, VSWIR

## 1. MOTIVATION FOR MONOLITHIC CATADIOPTRIC OBJECTIVES IN SPACE OPTICS

CubeSats have become a credible platform for Earth observation and small-aperture astronomy, but their small form factor constrains the maximum spatial resolution attainable by an imaging payload. Given an orbital altitude H, sensor pixel pitch p, a focal length of $f$ limits the spatial resolution to $\Delta x = p \cdot H/f$. The pixel pitch is typically 5µm for visible-range (VIS) and 15µm for short-wave infra-red (SWIR) camera sensors, so a target spatial resolution of 25m at a low Earth orbit (LEO) altitude of 500km requires f = 100mm and f = 300mm, respectively.

The physical length of refractive objective lenses is typically commensurate to their focal length, while their diameter is driven by maintaining the diffraction limit. Sizing the Airy disk diameter, given by $D_{Airy} = 2.44 \cdot \lambda \cdot F/\#$, to be within the above pixel pitch requires the objective to be roughly F/3 or faster, which in turn requires an aperture diameter of Ø33mm for VIS and Ø100mm for SWIR imagers. For hyperspectral imaging payloads in particular, fast collection optics are critical to maintain adequate signal-to-noise since the collected flux is dispersed among many spectral channels. Given that dimensions of CubeSat payloads are often constrained to one or two "U", or units of 10cm × 10cm × 10cm, it becomes challenging and at times impossible to achieve high spatial resolution imaging with standard refractive objective lenses. Because well-corrected refractive objectives are comprised of multiple lens elements, they also tend to be heavy and susceptible to mechanical shock. A robust optomechanical housing is required to keep launch vibrations from jolting individual lenses out of position, which further adds to the mass of the individual glass lenses.

* iliya.shofman@mail.utoronto.ca | ss-outreach@utat.ca | utat.ca/space-systems

Folding the optical path is required to achieve long focal lengths in a compact form factor. Classic examples include Cassegrain or Ritchey-Chrétien telescopes, which have been widely adopted in astronomy, some commercial Earth observation nanosatellites, and famously the Hubble Space Telescope. Reflective optics are inherently lightweight and achromatic, and figuring their surfaces into conic aspheres corrects spherical aberration and permits fast focal ratios without degrading on-axis image quality. Ritchey-Chrétien configurations correct coma alongside spherical aberration, yielding a wide well-corrected field of view.

The biggest challenges with realizing Cassegrain-like reflective objectives come from their sensitivity to misalignment and thermo-elastic deformation of the optomechanical structure. Reflective surfaces impose a larger wavefront error penalty of a comparable transmissive surface (a figure error $\delta$ yields $2\delta$ of wavefront error on reflection, versus $(n-1)\,\delta \approx 0.5\,\delta$ in transmission for glass with $n \approx 1.5$), and therefore demand tighter tolerances on surface form error. With only two surfaces carrying the entire optical power, a Cassegrain-like system also has far less freedom to compensate figure or alignment errors across elements than a multi-element refractive objective; the dominant challenge is consequently centering and angling the secondary precisely with respect to the primary. This in turn places strict requirements on the optomechanical housing to be very robust to mechanical shock and thermo-elastic deformation. Moreover, thermal expansion of the optomechanical housing produces a despace between primary and secondary mirror, which is then amplified by the system's longitudinal magnification, $m^2$. For example, at a modest secondary magnification of $m = 4$, a mere 10 µm of despace can displace focus by roughly 160 µm. Accurately modelling the performance of reflective objectives over the cyclic orbital temperature range requires structural-thermal-optical performance (STOP) analysis – a practice that is routine on flagship missions but generally beyond the scope of a CubeSat program.

Monolithic catadioptric lenses combine mirror and lens elements into a single, unified optical block rather than separate assembled components, which eliminates many of the alignment and thermal-related difficulties of reflecting objectives. Both mirrors, together with the refracting entrance and exit surfaces, are polished into the faces of a single blank so their separations are fixed at fabrication and cannot subsequently drift or be disturbed by launch. This intrinsic stability makes the monolithic form especially well-suited to compact long-focal-length objectives, where the folded catadioptric path packs a long effective focal length into a short physical envelope, and where any post-fabrication misalignment would otherwise be magnified into significant image degradation.

Designs in this lens class have been demonstrated recently, including Schmidt–Cassegrain monoliths [1], freeform-surface variants that fold off-axis to remove the central obstruction and shrink the footprint [2], and achromatized cemented-doublet catadioptric designs [3]. In the thermal IR wavelengths, single-point diamond turning (SPDT) has been leveraged to machine the optical surfaces and mechanical datum features in the same blank, enabling the objective to be seamlessly integrated within the rest of the system with high accuracy [4]. Commercial development of this lens class for space applications is also gaining momentum [5,6]. A parallel lineage of annular folded, gradient-index and metasurface architectures has developed largely in the academic literature [7–10].

Monolithic catadioptric lenses are typically made from fused silica, chosen for its low expansion coefficient, broad transmission band and excellent polishability. Since fused silica is a low-dispersion medium (Abbe number $v_d \approx 67.8$) and the reflecting surfaces can be designed to carry most of the optical work, the lens can be nearly achromatic across a broad spectral band. Fused silica is additionally well suited to the space environment, offering excellent resistance to radiation-induced degradation that compromises many conventional optical glasses [11]. Because fused silica has such a low coefficient of thermal expansion (CTE) – 0.55 ppm/K, against 7.7 ppm/K for BK7 – the monolithic lens preserves its surface form across a wide temperature range. Since the lens is fabricated from a single material, its physical length and its surface curvatures scale together under thermal expansion and contraction. To first order, this manifests as a change in focal length rather than a defocus. On the other hand, the thermo-optic coefficient (dn/dT) of fused silica is not itself small (10 ppm/K, versus 2.2 ppm/K for BK7), yet it has little bearing on performance. Because the optical power resides predominantly in the reflecting surfaces, and the refracting surfaces being correspondingly weak, the dn/dT contribution is likewise weak. Fused silica therefore renders the monolithic catadioptric lens remarkably and passively athermal, an advantage shared by neither purely refractive nor purely reflective objectives.

Our development of a monolithic catadioptric lens was motivated by CubeSat pushbroom hyperspectral imaging systems [12,13], which demand a compact, fast, and rugged objective lens in the VNIR and/or SWIR wavelength ranges. In a pushbroom hyperspectral imaging system, light from the objective lens is coupled into the slit. The slit serves as the entrance pupil to a spectrograph, which in turn disperses light perpendicular to the long axis of the slit and images it onto the camera sensor. The slit width dictates along-track spatial resolution and is often commensurate to (a multiple of) the pixel pitch. Maintaining in-focus coupling of light from the objective into the slit is a major optomechanical challenge. Consider, for example, an F/3 objective feeding a 15µm slit. A 45µm defocus doubles the diameter of the beam at the slit plane, and as a consequence only ~60% of the beam is transmitted. Holding a tolerance of a few tens of microns across an orbital thermal cycle and through launch vibration is demanding yet is critical to maintaining good light throughput - this is among the tightest interfaces in the instrument.

We propose to remove the difficulty entirely by integrating the slit onto the back surface of the monolithic lens structure. If the exit surface of a monolithic catadioptric objective is made flat and the design is optimized so that the focal plane lies coincident with it, the slit can be fabricated directly onto the focal plane with a lithographic process. In contrast to precision foil slits and adjustable mechanical slits, lithographically patterned and etched slits have highly accurate dimensional accuracy and mechanical robustness. The slit's alignment to powered optical surfaces is fixed during fabrication and is not susceptible to differential CTE between lens and barrel materials as is the case with traditional lenses. Moreover, coupling can be optimized by slightly adjusting the surface form of a preceding powered optical surface. The result is a single monolithic element that merges imaging optics and field-defining slit into a self-aligned, mechanically robust, passively athermal front end for a pushbroom spectrometer.

The remainder of this paper is organized as follows. Section 2 surveys the monolithic catadioptric design space, spanning Schmidt-Cassegrain derivatives, annular folded lenses, and gradient-index, diffractive, and metasurface variants. We focus our attention on designs in the Schmidt-Cassegrain lineage, which we compare by their wavelength range, focal length, F-number, and imaging performance. Section 3 presents our monolithic catadioptric objective with an integrated slit. We set the lens requirements, then develop three variants at f=100mm and F/3 over 0.4µm-1.7µm and ±3° of field, constraining asphere departure so that every surface can be certified on a standard Fizeau interferometer. For the preferred variant we show that a lateral singlet slide and a boss trim recover near-nominal as-built performance.

## 2. REVIEW OF MONOLITHIC CATADIOPTRIC LENS DESIGN LITERATURE

This section reviews the monolithic catadioptric lens literature. Section 2.1 surveys recent folded architectures - Schmidt-Cassegrain derivatives, annular folded, and gradient-index designs - alongside commercial and patented systems. Section 2.2 discusses the optical design principles governing these lenses, and Section 2.3 examines fabrication challenges and their suitability for nanosatellite payloads, which serves as the motivating application for this study.

## 2.1 Review of Monolithic Catadioptric Folded Lenses

This section reviews the published literature, patent record, and commercial offerings of monolithic catadioptric lenses. We trace the historical development of the design form, survey modern implementations in the Schmidt-Cassegrain lineage, and summarize the parallel lineage of annular folded, gradient-index, and diffractive designs. Representative examples of each class are compared in Table 1 by focal length, F-number, number of folds, wavelength range, and imaging performance. Several representative examples of various monolithic lens forms are collected in Figure 1. To the authors' best knowledge, no comprehensive review of this lens class has been published.

### 2.1.1 Historical development of monolithic catadioptric lenses

Catadioptric optical components can be traced back to the Mangin mirror, originally conceived as a searchlight reflector in 1876. A single glass element combines a refracting front surface with a silvered rear reflecting surface so that the reflector's spherical aberration is corrected within one piece of glass. The first proposal to consolidate an entire two-mirror telescope into a solid block came from Maksutov, who in the mid-twentieth century suggested that his meniscus-corrector system could be made more rugged and stable by fabricating it as a glass mono-bloc [14]. The concept went unrealized owing to its fabrication difficulty: four spherical surfaces had to be centered on a common axis, the central block thickness held to a tight tolerance, and the sag of every surface accounted for in design and fabrication. The first monolithic two-mirror telescope to be built and published appeared in 1963; however, its all-spherical construction limited the useful focal ratio and left the design poorly corrected [15].

The solid catadioptric form subsequently progressed largely through the patent literature. Richards (1991) described a hybrid element with a reflective outer annular zone surrounding a refractive core - each guiding light through a separate sequence of surfaces - conceived to increase the numerical aperture of focusing optics [16]. Cameron and Sturiale described a "solid catadioptric lens" (1998) consisting of a planar input surface, concave primary and convex secondary mirrors, and a spherical exit surface, without specifying the optical performance achieved and surface prescriptions [17]. Maresse described an "ultra-compact mono-bloc catadioptric imaging lens" (2008), which revisited the Maksutov-type construction with two refractive and two reflective surfaces, with all aspheric surfaces in the preferred embodiment [18]. Beyond compact tele-photo imaging, the lens form principle was proposed for use in document scanning [19], deep-UV lithography [20], and folded-path sensing and metrology [21,22].

A recurring theme of this early period is that manufacturing capability, rather than optical design, limited what could be realized [14,15]. The enabling advances arrived late in the twentieth century: single-point diamond turning, CNC and deterministic sub-aperture polishing, magnetorheological finishing and computerized polishing, along with freeform metrology such as profilometry and deflectometry. Together these made aspheric and freeform surfaces practical to fabricate and verify, removing the focal-ratio and aberration limits of the all-spherical era and enabling the modern implementations surveyed next.

### 2.1.2 Recent development of monolithic catadioptric lenses

The earliest modern fabricated monolithic catadioptric lenses were based on the Schmidt-Cassegrain design and leveraged aspheric surfaces to produce good imaging quality. ter Horst published monolithic Schmidt-Cassegrain telescopes in the 1990s [23], consolidating the refractive and reflective surfaces, together with internal baffles, into a single fused silica block [1]. These "Tiny Telescopes" have since been developed toward nanosatellite Earth observation, acquiring flight heritage when a f=125 mm F/5 solid Schmidt-Cassegrain flew aboard the OreSat student CubeSat. Refinements including an aspherical entrance surface and use of the exit surface for field correction have brought the design to diffraction-limited image quality [6], now offered commercially by Tiny Telescope B.V. [24].

Galan et al. described a solid catadioptric lens based on a Ritchey-Chrétien design [3]. In the infrared, the authors demonstrated two diffraction-limited variants in the 8-12µm and in the 3-5µm wavelength range, using a substrate of a single glass material. Popular infrared glass materials, including germanium (Ge), gallium arsenide (GaAs), zinc selenide (ZnSe) and zinc sulfide (ZnS), have a low dispersion in the mid-wave and long-wave infrared which results in little chromatic aberration. Since dispersion of glass is substantially larger in the visible range, Galan et al. proposed a cemented doublet design with aspheric reflecting and refracting surfaces to mitigate chromatic aberration. This design achieved a telephoto ratio of 0.6 with a field of view of 4.2° by 4.2° at F/4.3 with nearly diffraction-limited imaging over 0.4-1.0µm wavelength range.

The largest commercial development effort originates at Lawrence Livermore National Laboratory (LLNL), whose "MonoTele" concept was inspired by the monolithic primary-tertiary mirror LLNL designed for the Vera C. Rubin Observatory [25]. Patented variants include a four-surface widefield form with two aspheric refractive and two aspheric mirror surfaces [26] and lightweighted air- and aerogel-core constructions [27], with physical realizations having apertures of 8.5cm and 18cm [28]. Substantial flight heritage was demonstrated by incorporating these lenses into the GEOStare1 (2018) and GEOStare2 (2021) imaging payloads, which demonstrated suitability for space domain awareness, tracking satellites and orbital debris, as well as astronomy and Earth observation [29]. The technology is being commercialized by Starris: Optimax Space Systems.

A further family of designs removes the central obscuration entirely by folding the system off-axis with freeform surfaces. Optimax, in collaboration with NASA Goddard Space Flight Center, manufactured three such monoliths as proofs of concept: a two-freeform-mirror solid glass prism sized for a 1U CubeSat (~1.3 kg); a three-freeform, diffraction-limited solid monolith (~2.3 kg); and a lightweighted two-freeform-mirror variant in which the rays no longer travel through glass (~0.42 kg) [30]. All three were designed for a ±4.3° by ±1.4° field of view at F/3.4, with mirror surfaces described by XY polynomials [2]. Finally, an afocal branch of monolithic telescopes places confocal paraboloids on the front and back faces of a single transmissive block [31, 32]. The arrangement performs collimated-beam reduction, is permanently aligned, and corrects all primary Seidel and chromatic aberrations [31]. Proposed applications include 20-50 cm class laser beam expanders for free-space optical communication.

### 2.1.3 Annular folded designs

A subset of monolithic catadioptric lenses, sometimes referred to as annular folded lenses (AFLs), fold the optical path more than twice to achieve a dramatic reduction in the lens physical length relative to its effective focal length. Several demonstrations of this lens with four and eight reflections between concentric annular mirror zones on the front and back faces of a single substrate have been made in the literature [7, 33]. The advantage of this approach is an extremely thin package with a large outer aperture. On the other hand, the large central obscuration depresses mid-spatial-frequency MTF contrast, and the annular zone-boundary discontinuities make this lens very challenging to fabricate compared to the two-reflection monoliths.

Tremblay et al. presented the multiple-fold approach together with an eightfold prototype camera in $CaF_2$, with a 35 mm effective focal length at 0.7 NA, 60 mm outer diameter and 90% obscuration folded into just 5 mm of total thickness. A follow up study by the same group demonstrated a four-reflection imager achieving F/1.15 and 18.6 mm focal length in a 5.5 mm track, with a 17° field of view over 1.92 megapixels of a 3 um-pixel color sensor [33]. This lens was comprised of two plano-aspheric $CaF_2$ lens elements machined with single-point diamond turning, with a small adjustable air gap between the two plano surfaces for refocusing the lens to different working distances.

### 2.1.4 GRIN and DOE designs

Aside from the cemented-doublet approach in [3], most monolithic catadioptric lenses do not correct for chromatic aberration and lateral color aberration, because the reflecting design is mostly achromatic. This is true when the refracting surfaces have low optical power; however, constraints on the curvature of these surfaces can affect imaging performance and reduce the lens field of view.

Two classes of unconventional optics have been applied to recover broadband performance. Freeform gradient-index (F-GRIN) media turn the substrate volume itself into a distributed corrector. By allowing the dispersion of the index gradient to be tailored independently of its spatial profile, color aberration can be compensated without adding surfaces or different materials as in traditional refracting objectives. Lippman et al. showed that F-GRIN monolithic AFLs achieve both higher monochromatic performance and full-visible color correction, reporting a diffraction-limited f = 100 mm, F/1.5 design with an impressive 30% contrast at 200 lp/mm across the field and a telephoto ratio of 0.56 [9].

Diffractive optical elements (DOEs) exploit a complementary property: the effective dispersion of diffraction is strongly negative, so a single microstructured surface can achromatize the refractive surfaces. An achromatic AFL with a reflective-diffractive element achieved visible-band MTF above 0.25 at 111 cycles/mm with diffraction efficiency accounted for [8]; a multilayer-DOE variant extended the approach across the visible and near-IR (0.45-1.1 µm) at a total-length-to-focal-length ratio of only 0.332 [34]; and a refractive-metasurface hybrid carried it into the mid-infrared [10]. Diffractive optical elements come with a well-known constraint as the diffraction efficiency falls away from the design wavelength and incidence angle, resulting in a reduction in MTF. On the other hand, the polymer and molded glass substrates used in the F-GRIN and DOE approaches have CTE and thermo-optic coefficients up to one to two orders of magnitude larger than fused silica and may perform poorly over a wide temperature range.

### 2.1.5 Comparison of monolithic catadioptric lens specifications

Table 1 compares representative examples: three from the solid catadioptric lineage that is the focus of this paper, two annular folded prototypes, and one design each from the GRIN and DOE families. Solid Schmidt-Cassegrain-derived monoliths dominate the fabricated and flight-proven examples, operating at moderate focal ratios over narrow, well-corrected fields. Annular folded lenses trade obscuration and fabrication complexity for very fast focal ratios in millimeter-scale tracks. The GRIN and DOE variants remain design studies and have not been extensively validated through fabrication. Monolithic catadioptric lenses with just two reflecting surfaces are challenged to reduce off-axis fields and achieve fast focal ratios. In comparison, the annular-folded lenses have many more surfaces and thus degrees of freedom to correct off-axis aberrations, though at the expense of a larger obscuration ratio.

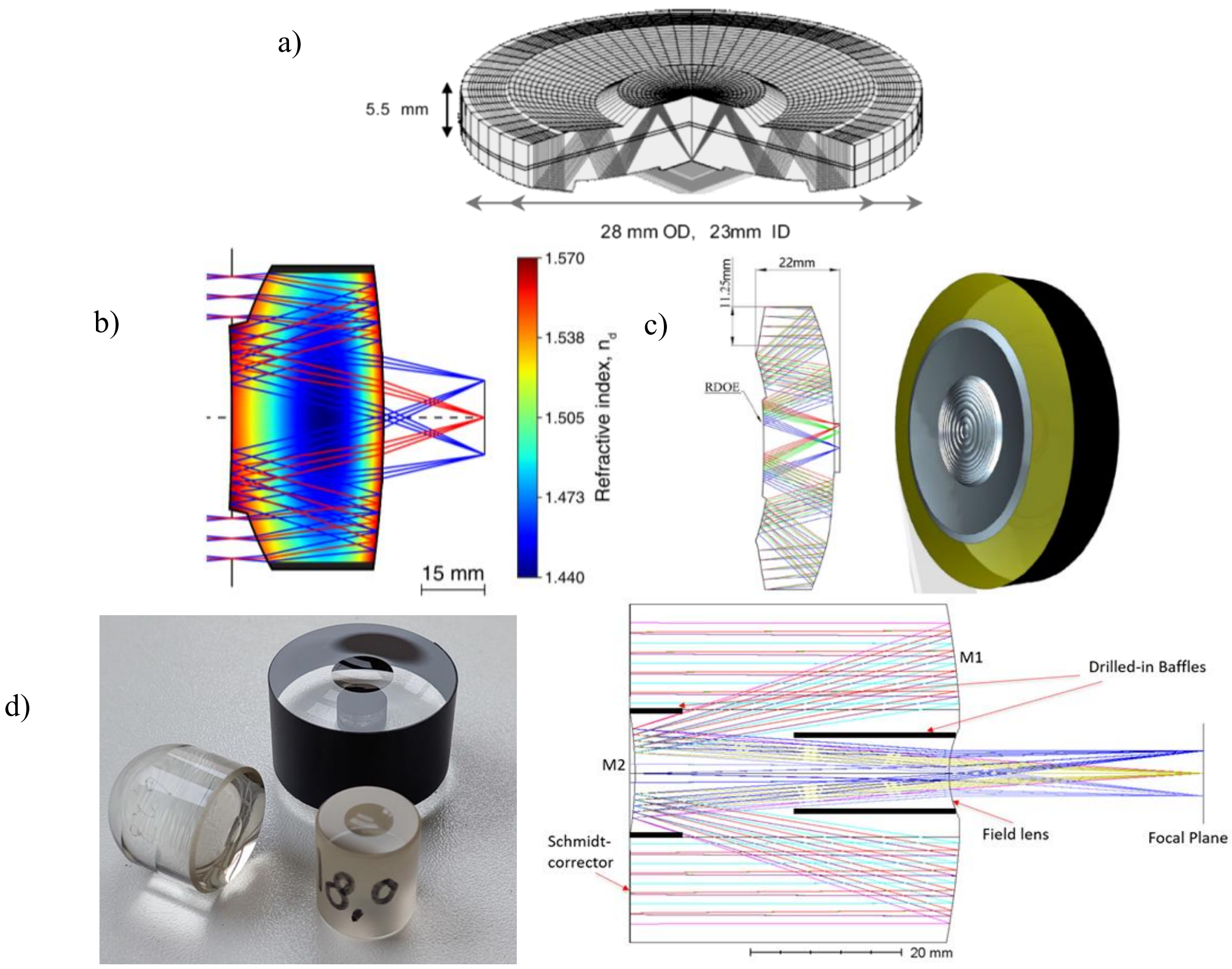

**Figure 1**. Monolithic catadioptric lens design forms; a) annular-folded four-reflection lens [33], b) F-GRIN lens [9], c) monolithic lens with diffractive optical elements [34], d) Schmidt-Cassegrain-like two-reflection monolithic lens [1].

**Table 1**. Comparison of representative monolithic catadioptric designs. n/r = not reported. L/f = telephoto ratio.

| **Design** | **Lens type / class, maturity level** | **Focal length** | **F/#** | **# folds** | **Performance Notes** | **Wavelength range** |
|---|---|---|---|---|---|---|
| Tiny Telescope (OreSat) [1,6,24] | Solid Schmidt-Cassegrain; commercial, flight heritage | 125 mm | F/5 | 2 | Near-diffraction-limited; 2.8° FOV; 28 mm total length; 29 g | VIS |
| LLNL MonoTele / Starris [25,28,29] | Solid catadioptric slab; commercial, flight heritage | n/r | n/r | 2 | Diffraction-limited; apertures 25-185 mm; 18.5 cm variant | VIS, SWIR variants |
| SoCatS, Galan et al. [3] | Maksutov solid + cemented doublet; design study | 110 mm, estimated | F/4.3 | 2 | Near-diffraction-limited over 4.2 x 4.2° FOV | VIS, MWIR, LWIR variants |
| Freeform monolith, Blalock et al. [2,30] | Off-axis freeform monolith; fabricated prototypes | 200 mm, estimated | F/3.4 | 2-3 | Diffraction-limited, three-mirror variant; ±4.3° x ±1.4° FOV; 0.42-2.3 kg | VIS |
| Eightfold AFL, Tremblay et al. [7] | Annular folded ($CaF_2$); fabricated prototype | 35 mm | 0.7 NA | 8 | 60mm outer diameter 90% obscured results in 27mm “effective aperture” in 5 mm track; imaging demonstrated | VIS |
| Four-reflection AFL [33] | Annular folded ($CaF_2$); fabricated prototype | 18.6 mm | F/1.15 | 4 | 17° FOV over 1.92 Mpx (3 µm pixels); 5.5 mm track | VIS |
| F-GRIN AFL, Lippman et al. [9] | Annular folded + freeform GRIN; design study | 100 mm | F/1.5 | 2 | Diffraction-limited across FOV; ~30% MTF at 200 lp/mm; L/f = 0.56, FOV ±5° | VIS |
| MLDOE AFL, Piao et al. [34] | Annular folded + multilayer DOE; design study | 50 mm | F/1.2 | 4 | MTF > 0.26 at 166 lp/mm incl. DOE diffraction efficiency; L/f = 0.33, FOV 7.7° | VIS-NIR |

## 2.2 Design Principles

The lenses surveyed in Section 2.1 differ widely in geometry and application, yet their behavior is governed by a common set of design principles, inherited from the classical two-mirror telescope and modified by the glass medium. This section collects these principles from the literature and our own modelling work on the Schmidt-Cassegrain form - a homogeneous glass block bounded by two refracting and two reflecting surfaces.

### 2.2.1 Third-order aberrations and the role of aspheric surfaces

The imaging performance of a two-mirror telescope is limited by the five third-order (Seidel) aberrations: spherical aberration, coma, astigmatism, field curvature, and distortion. For a system of focal ratio $N = f/D$ imaging a semi-field angle $\theta$, the angular blur of the first three scales characteristically with aperture and field [35, 36]: spherical aberration as $N^{-3}$, independent of field; coma linearly with field, as $\theta/N^2$; and astigmatism quadratically with field, as $\theta^2/N$. Comparing against the fixed angular diffraction limit, $2.44\lambda/D$, aberration scaling rules can be used to estimate the usable field. Once spherical aberration is corrected, the semi-field over which coma stays below the diffraction limit scales as $\lambda N^3/f$. Once coma aberration is corrected, the semi-field is limited by astigmatism and scales as $N\sqrt{\lambda/f}$.

Each powered mirror, figured as a conic or higher-order asphere, provides the freedom to cancel one third-order aberration: a single parabolized mirror is free of spherical aberration; asphérizing both mirrors as in the Ritchey-Chrétien additionally removes coma for an aplanatic design; and a three-mirror anastigmat additionally corrects astigmatism [37]. The monolithic Schmidt-Cassegrain follows a similar principle: the aspheric Schmidt entrance face removes the spherical aberration of the spherical primary, the secondary is aspherized to control coma, and the exit face remains available for further field correction [35].

The aperture stop position is an additional aberration-control variable cleverly exploited in the Schmidt-Cassegrain design. The corrector and aperture stop are placed at the center of curvature of the spherical primary, so every chief ray passes through the center of curvature and each field point sees the mirror as though on axis. By symmetry the primary contributes no coma, astigmatism, or distortion; what remains is spherical aberration - removed by the corrector, and field curvature – resolved by making photographic film conform to a curved image plane. These powerful cancellations are the reason why classical Schmidt cameras covered wide fields of view of many degrees [35, 36]. A compact monolith cannot afford this geometry because the block is far shorter than the primary's radius of curvature. As a result, the stop sits well inside the center of curvature and coma and astigmatism return, growing with its displacement.

### 2.2.2 Central obscuration, field of view, and MTF

An on-axis folded lens necessarily images through a central obscuration, characterized by the ratio $\alpha$ of obscured to full aperture diameter. The obscuration costs throughput, reducing the collecting area by $1 - \alpha^2$, and reshapes the diffraction point-spread function, transferring energy from the Airy core into the surrounding rings. This has the effect of depressing MTF contrast at mid-spatial frequencies while leaving the incoherent cutoff frequency, $1/\lambda N$, unchanged [9]. Minimizing $\alpha$ is thus a driving design factor across the class, from heavily folded annular designs where $\alpha$ can reach 0.9 down to the modest obscurations of two-reflection monoliths where $\alpha$ is closer to 0.5 [7].

The obscuration is also coupled to the field of view. The secondary mirror must intercept the full beam footprint of every field point, and off-axis footprints walk laterally across the mirror in proportion to the field angle and the primary-secondary mirror separation. Widening the field therefore demands a larger secondary mirror, which raises $\alpha$; conversely, holding $\alpha$ fixed caps the unvignetted field. Field of view, obscuration, and compactness thus form a closed trade, which Galan et al. identify as the envelope of the solid catadioptric class, restricting practical fields to below roughly 5° × 5° and focal ratios to about F/4 and slower [3].

### 2.2.3 Thermal behavior of the fused-silica monolith

The monolithic form is athermal in alignment but not in refractive index. The low CTE of fused silica (0.55 ppm/K) ensures that surface separations and radii barely change with temperature and, as noted in Section 1, what motion remains is a self-similar rescaling of the entire lens element. However, the thermo-optic coefficient of fused silica, $dn/dT \approx 10$ ppm/K, is roughly twenty times its CTE. With light taking multiple passes through the thick fused silica block, the thermal error budget is therefore dominated not by the structure expanding but by the refractive index of the glass changing [38].

Two regimes must be distinguished. A uniform temperature change shifts the index uniformly, producing to first order a small focus shift, which is correctable by, for example, a CTE-selected detector standoff. A temperature gradient across the block is far more detrimental as the spatially varying index turns the block into a weak gradient-index lens, producing higher-order wavefront error in addition to defocus. The accumulated error is approximately OPD $\approx (\mathrm{d}n/\mathrm{d}T) \cdot L \cdot \Delta T$ over an in-glass path $L$. For the 100-200 mm folded paths typical of this class, a 0.1 K transverse gradient produces 100–200 nm of wavefront error – $\approx\lambda/10$ at SWIR wavelengths. Orbital solar loading can produce such gradients, and the low thermal conductivity of glass precludes temperature gradients from equalizing quickly across the bulk of the monolith [38].

Several mitigation strategies have been proposed [38]: engineering the thermal boundary conditions so the block stays isothermal; athermalizing with a second material whose thermo-optic behavior opposes fused silica's; and absorbing the residual focus shift mechanically. Reflection itself is governed by geometry alone, so the thermo-optic sensitivity arises from entrance and exit refractions and bulk propagation – unavoidable in a solid catadioptric lens.

### 2.2.4 Residual chromatic aberrations

The paraxial behavior of the solid form can be understood by decomposing it into equivalent thin lenses: entrance face to primary as a weak positive lens, primary-secondary pair as a second positive lens, and secondary to exit face as a negative lens [3]. Each element contributes axial color in proportion to its power over its dispersion as $\Sigma\, y^2\varphi/v$, with $y$ being the marginal ray height, $\varphi$ the power, and $v$ the Abbe number. Since the mirrors are achromatic and the refracting surfaces weak, the residual is small but not zero. Therefore, crown glasses with low-index, high-Abbe-number are routinely chosen for monolithic lens designs, which fused silica ($v_d \approx 67.8$) fits naturally. Where tighter correction is needed, a flint element of opposing power cancels axial color ($\Sigma\, \varphi/v \rightarrow 0$). This can be accomplished by splitting the block into a cemented crown-flint doublet for the visible [3] or inserting a corrective flint singlet or doublet that precedes the monolith [26].

Finally, these lenses exhibit a lateral color error that air-spaced reflective telescopes do not. The chief ray refracts at the flat or near-flat entrance face through a wavelength-dependent angle governed by Snell's law – and the angular dispersion accrues into a lateral image displacement over the long glass path, with no subsequent surface carrying opposing dispersion to cancel it. The magnitude is small, with Tremblay et al. reporting ~8 µm lateral color at full field in the visible range, and can be removed in post-processing for an imaging camera [7]. However, for instruments coupling light into a slit the tolerance is stricter: along the slit, lateral color is a wavelength-dependent magnification which results in keystone distortion in a hyperspectral data cube, and across the slit it makes the coupling efficiency wavelength-dependent even at best focus.

### 2.2.5 Further considerations unique to the monolithic form

Because imaging and stray-light paths share the same block, baffling must be built into the blank itself. Both ter Horst and the LLNL designs integrate baffles around the exit surface, fabricated by drilling cylindrical recesses into the block and blackening them with absorptive paint or an epoxy loaded with absorbing material [1,25].

Two further properties follow from immersing the optical path in glass. Refraction from air into glass at the near-flat entrance face compresses the field, mapping an external field angle $\theta$ to $\theta' \approx \theta/n$ inside the block – about 70% of its external value in fused silica. The reduced internal field is favorable because the mirrors generate less off-axis aberration. As a result, the optical speed of the solid Schmidt-Cassegrain lens can be $\sim n^2$ larger than for an air-filled equivalent [36]. The second property is less forgiving: reflecting surfaces must be figured more accurately. A figure error $\varepsilon$ on a transmissive face contributes only $(n - 1)\, \varepsilon \approx 0.45\varepsilon$ of wavefront error, whereas the same error on a glass-immersed mirror contributes $2n\varepsilon \approx 2.9\varepsilon$, roughly six times as much, and a factor of n worse than the same mirror in air. Steep or otherwise demanding aspheres are therefore better placed on the more error-tolerant refracting Schmidt surface, albeit at the cost of substantially larger surface sag [1].

## 2.3 Fabrication and Metrology Considerations

The monolithic catadioptric lens is a powerful lens form that provides long focal lengths in very compact form factors, has excellent mechanical robustness and athermal performance, and can offer near-diffraction-limited imaging over broad wavelength ranges with a modest field of view. On the other hand, this lens form is somewhat unconventional from a manufacturing standpoint, requiring special considerations for tolerancing and alignment, which we discuss in this section.

### 2.3.1 Fabrication heritage: diamond-turned monoliths in the thermal infrared

Monolithic optics first became practical in the thermal infrared, when advances in computer-controlled single-point diamond turning (SPDT) enabled fabricating optical surfaces and mechanical assembly structures on the same substrate [39]. This type of construction reduces many optomechanical tolerances down to machining accuracy while making the optical assembly more robust to mechanical shock. By integrating mounting flanges, interface flats, and alignment datums with the optics, surface-to-surface registration inherits micrometer-level machine accuracy instead of depending on assembly adjustment. The monolithic design reduces the number of inter-element fasteners and adhesive bond joints, whose loosening under vibration is a common source of alignment drift. While SPDT leaves turning marks that result in mid-spatial frequency (MSF) error, they are generally acceptable for applications in the 8-12 µm range; subsequent polishing can reduce MSF error and surface roughness to a level acceptable for SWIR and even visible-range optics.

On the other hand, not all materials are compatible with single-point diamond turning. SPDT is suited to non-ferrous metals such as aluminum and electroless nickel, to chalcogenide glasses and infrared crystals such as germanium and zinc sulfide, as well as calcium fluoride and certain optical polymers. What these materials share is that they are soft or ductile enough to shear cleanly at the microscopic cutting depth a diamond tool takes. Fused silica and other silicate glasses fall outside this set because their hardness drives machining into the brittle-fracture regime, leaving a fractured surface with substantial sub-surface damage and wearing the diamond cutting tool rapidly. These materials are therefore figured by grinding and polishing or by CNC sub-aperture techniques instead.

Single-point diamond turning has been leveraged for the fabrication of several monolithic catadioptric lenses. Tremblay et al. fabricated an eight-fold visible annular folded lens with SPDT and subsequent magnetorheological finishing (MRF) on a calcium fluoride substrate [7]. Since multi-zone annular mirror surfaces can be machined without re-chucking the optic, zone decenter tolerances on the few-micrometer level can be obtained just from machine precision itself [40].

### 2.3.2 Fabrication sequence and metrology for a fused silica monolith

A monolithic lens made of fused silica would follow a fabrication sequence consisting of CNC generation, lapping, sub-aperture polishing, and deterministic figure correction. This subsection walks through the fabrication sequence and the tolerances typical of each stage and summarizes them in Table 2 below.

Generation of the powered surfaces follows, with best-fit spheres ground into each face using CNC tooling. Radius can be tracked in process using a spherometer which can resolve sag to about 1 µm, corresponding to radius accuracy at the 0.1% level for surfaces of this lens class. Final radius can be verified to 0.01–0.05% accuracy on a Fizeau interferometer with a transmission sphere whose cat's-eye-to-confocal travel is tracked by a displacement-measuring interferometer. The CNC can cut the base sphere of each annular surface zone on a single face of the monolith in a single chucking, so their relative registration inherits machine accuracy rather than assembly accuracy. Tool setting and thermal drift limit the accuracy of zone-to-zone decenter to 2–5 µm, axial step between zone vertices to ±2–5 µm, and tilt between zone axes to 0.01° [40].

Figuring the opposite face requires a re-chuck, and hence the monolith is re-centered optically with an autocollimator. Electronic autocollimators measure to 0.5–1 arcsec, which corresponds to ~0.25µm of lateral offset between the faces across a 50-mm-thick block. In practice, accuracy is limited by chuck adjustment mechanics to 1–3 µm of decenter and 5–20 arcsec of tilt [2]. However, it is challenging to achieve accurate center thickness specs, with the typical precision-level tolerance grade being ±0.05mm [41]. The limitation comes from the nature of the grinding process, which removes material at a fast rate that depends on polishing pad wear, slurry condition, pressure, and temperature. Since material is removed until the correct figure is achieved, it is difficult to simultaneously aim for an exact thickness target. On the other hand, center thickness can be measured to very high accuracy with low-coherence interferometry to the 0.2 µm level. The practical strategy for this lens class is therefore to measure center thickness and compensate for it, as discussed later.

After the base sphere has been figured, mild aspheres are imparted by sub-aperture bonnet or fluid-jet polishing, and where required, magneto-rheological finishing. CNC sub-aperture polishing converges surface form accuracy to ≈λ/4 PV but leaves mid-spatial-frequency ripple while MRF finishes surfaces to ≈λ/20 PV [44]. Accuracy of aspheric surfaces can be verified using a standard Fizeau interferometer in a non-null test against a transmission sphere, so long as the slope departure from the best-fit-sphere is not too large. Fringes are resolvable up to the Nyquist frequency of two pixels per fringe or a surface slope of λ/4 per pixel. Quantitative measurements can be made with a frequency of one fringe per six pixels or lower. A representative mild asphere with a peak departure slope of 1.3 µm/mm produces four fringes per millimeter at 633 nm. Meanwhile, a 1200-pixel interferometer camera samples ~35 µm per pixel when zoomed to a Ø40 mm clear aperture, giving seven pixels per fringe – inside the measurable bound. Aspheres with a few micrometers of departure are therefore certifiable on a standard Fizeau interferometer, provided the departure slope remains within ≈1 µm/mm at the tested magnification.

For higher departure aspheres, more sophisticated metrology tools are required. For example, Zygo Verifire Asphere+ workstations translate the optic and computationally stitch together locally-nulled annular zones recorded on a Fizeau interferometer. Such instruments accommodate up to ~800 µm of departure from the best-fit sphere, provided that the surface lies within ~10 µm of the nominal prescription, and the part must be rotationally symmetric with a measurable vertex at its center [42].

Multi-zone surfaces of the monolithic lens cannot be measured using this device because the vertex of any annular zones is carried by the central zone which may be out of range. Instead, computer-generated hologram (CGH) nulling tests are typically employed and can yield 10nm RMS accuracy, though each hologram is expensive and specific to a single prescription. Alternatively, contact profilometry can be used to measure surface form independent of slope or sag but only offers 50–100 nm of form accuracy, which is adequate at SWIR wavelengths yet marginal for certifying near-diffraction-limited figure in the visible. Keeping the aspheres mild enough for non-null Fizeau testing is therefore a metrology-driven constraint, and one the present design adopts and develops in Section 3.

**Table 2**. Typical metrology techniques and their respective accuracy, for fused silica monolithic catadioptric lens [41].

| **Quantity** | **Instrument class** | **Typical accuracy** |
|---|---|---|
| Flatness of blank faces | Fizeau interferometer + transmission flat | λ/20 PV (reference-limited); sub-nm repeatability |
| Parallelism (wedge) of blank | Fizeau / dual autocollimator | 0.1–0.5 arcsec |
| Center thickness & spacing | Low-coherence interferometry | 0.15–0.5 µm |
| Surface-axis decenter | Autocollimating centration station | 0.1–0.2 µm |
| Radius of curvature | Spherometer | 0.05-0.1%, depending on curvature & diameter |
| Radius of curvature | DMI-tracked interferometric radius bench | ~$10^{-4}$ relative |
| Sphere figure | Fizeau + transmission sphere | ~λ/20 PV |
| Mild asphere figure (≤ few µm departure) | Non-null Fizeau (zoomed, best-compensation defocus) | Slope-limited; quantitative only below ~1/3 of camera Nyquist |
| Strong asphere figure – null test | Fizeau + computer-generated hologram (CGH) | ~5–10 nm RMS; pattern placement ~0.1 µm; per-prescription artifact |
| Strong asphere figure – stitching | Annular scan asphere interferometry workstation | 1–5 nm RMS repeatability; ~10 nm-class accuracy |
| Strong asphere figure – profilometry | Contact stylus or optical point-probe profilometer | 50–100 nm form; independent of aspheric departure |

### 2.3.3 Stray-light treatment and photolithographic definition of the entrance slit

After the lens surfaces have been figured, reflective zones can be broad-band mirror-coated using one of several techniques. Protected aluminum or silver is applied by evaporation or magnetron sputtering over a thin adhesion layer; annular coverage is obtained with mechanical shadow masks. Non-reflecting and non-refracting surfaces can be coated with black paint to mitigate stray light.

More meaningful stray-light suppression can be added with a core-drilled baffle bore that is coated with absorptive paint or filled with blackened epoxy [1, 25]. While the bore cut need not be as dimensionally precise as the rest of the monolithic lens, there is a large risk of edge chipping and subsurface cracking during cutting. It is crucial to feed the bore cutting tool gently and flush the blank with water during cutting. To mitigate the risk of critical damage from the bore, the cut can be made before investing significant time into figuring and inspecting the complex optical surfaces of the monolith.

As discussed in Section 1, we propose integrating a slit onto the back flat-polished surface of the monolith. Photolithographic techniques have been used to produce optical slits with high precision and robustness and have gained space-heritage on satellite hyperspectral imaging payloads [12, 43]. To fabricate the slit, an opaque chromium film is deposited over the substrate and coated with photoresist. The slit pattern is exposed by laser illumination, developed to open the resist, and transferred into the chromium by wet etching. Chromium is a standard mask metal because it adheres strongly to bare glass with no underlayer and a thin film of only ~100–120 nm is pinhole-free and optically dense. The chromium film is hard and durable, and it has a mature selective wet-etch chemistry that leaves the glass untouched. Bare chromium reflects ~55–65% in the visible range, so an absorptive chromium-oxide layer can be added between the glass and the metallic chromium to suppress stray light that would otherwise reflect off the slit mask and back into the instrument. This stacked mask is standard and is widely used commercially, though it introduces some complexity on the etching step.

The achievable precision with a photo-mask slit is very high: width tolerance is sub-micron, edge roughness is on the tens-of-nanometer scale, and the placement relative to fiducials on the same surface is very accurate. Conventional air slits in etched or electroformed metal foil, for comparison, have width tolerances of a few microns and micron-scale edge roughness. Their finite-thickness walls can also vignette fast beams. More importantly, unlike the proposed integrated slit, air slits must be mechanically mounted, and maintaining focus becomes a liability in rugged operating environments.

## 3. DESIGN AND ANALYSIS OF MONOLITHIC LENS WITH INTEGRATED SLIT

The development of our monolithic catadioptric lens with an integrated slit was motivated by the FINCH satellite mission, a student-built 3U CubeSat with a SWIR hyperspectral imaging payload. A long focal-length lens is needed to attain a scientifically actionable spatial resolution within the limited nanosatellite form factor. Simultaneously, since light is dispersed into many wavelength channels, the objective lens must be optically fast. Considering also requirements on mechanical ruggedness to withstand launch vibrations and thermal stability against on-orbit temperature fluctuations, the monolithic catadioptric lens form is the most suitable for this application. Indeed, other researchers have explored using a monolithic catadioptric lens in a nanosatellite hyperspectral imaging payload [12].

In a pushbroom hyperspectral imager, the objective lens is coupled to a spectrograph through a slit, yet maintaining the slit in the focus of an optically fast beam is a significant optomechanical challenge. We can eliminate this by designing the monolithic lens such that its image plane coincides with its polished-flat back surface, and then lithographically etching a slit onto that surface. By choosing fused silica as the lens material, we can leverage its low CTE to maintain focus at the slit over a wide range of temperatures. To the authors' best knowledge, this idea has not yet been proposed or developed in publicly available literature.

Since the monolithic lens form is a non-standard lens form and is challenging to fabricate, we added constraints to relax manufacturing tolerances and simplify the metrology of the lens. We constrained the sag and slope departure on aspheric surfaces so they can be readily inspected with a standard Fizeau interferometer, as opposed to sophisticated high-departure asphere metrology stations or computer-generated holograms. We also propose several compensation strategies that can restore nominal performance of the as-built lens with fabrication tolerances, while retaining the mechanical benefits of a monolithic lens.

### 3.1 Preliminary Lens Requirements and System Design Considerations

In this work, we developed several variants of the monolithic catadioptric lens form, sketched in Figure 2. All variants are designed for an effective focal length of f=100mm and an aperture of F/3. A mechanical diameter of 50mm allows for some clear aperture margin and access for mechanical interfaces. Since light throughput is critical to a hyperspectral imaging application, we opted for a two-mirror design with ≈50% diameter obscuration. While annular-folded lenses with four, six, and eight mirrors have been demonstrated to offer wider well-corrected field of view, those lenses have large obscuration ratios and hence have a low light collection ability. The two-fold design places the thickness of the monolith around 50mm, which yields a telephoto ratio of ≈0.5.

We optimized the lens over λ=0.4μm-1.7μm for field angles up to ±3°, which corresponds to a ground swath width of 50km in low Earth orbit. We found that the central obscuration had to be expanded slightly to prevent clipping the off-axis beams, and as a consequence only ≈70% of incident light rays are transmitted to the image plane. Optimizing lens performance simultaneously across the VSWIR spectrum (0.4-1.7μm) leverages the inherently good achromatic performance of this lens form that can be used in broadband hyperspectral imaging or be tailored specifically for VNIR or SWIR at a subsequent stage. Since the dispersion of fused silica begins to increase for wavelengths approaching the UV, the optimization failed to meet diffraction-limited performance at λ=0.4μm by a small margin.

Our lens design is well adapted for modern SWIR and VSWIR sensors. For example, the Sony IMX990 image sensor has 1293 (H) by 1032 (V) pixels with 5μm pixel pitch; its 6.5mm horizontal side would be subtended by about ±2° field of view by our f=100mm lens. The full ±3° field of view would span 10.5mm on the sensor plane. We note that the 5μm pixel size is slightly larger than the diffraction-limited spot size at λ=0.4μm for F/3, while the Airy disk at λ=1.7μm would be covered within 3×3 pixels. Hence, a 15μm slit width is a reasonable value for a pushbroom hyperspectral imager using this sensor.

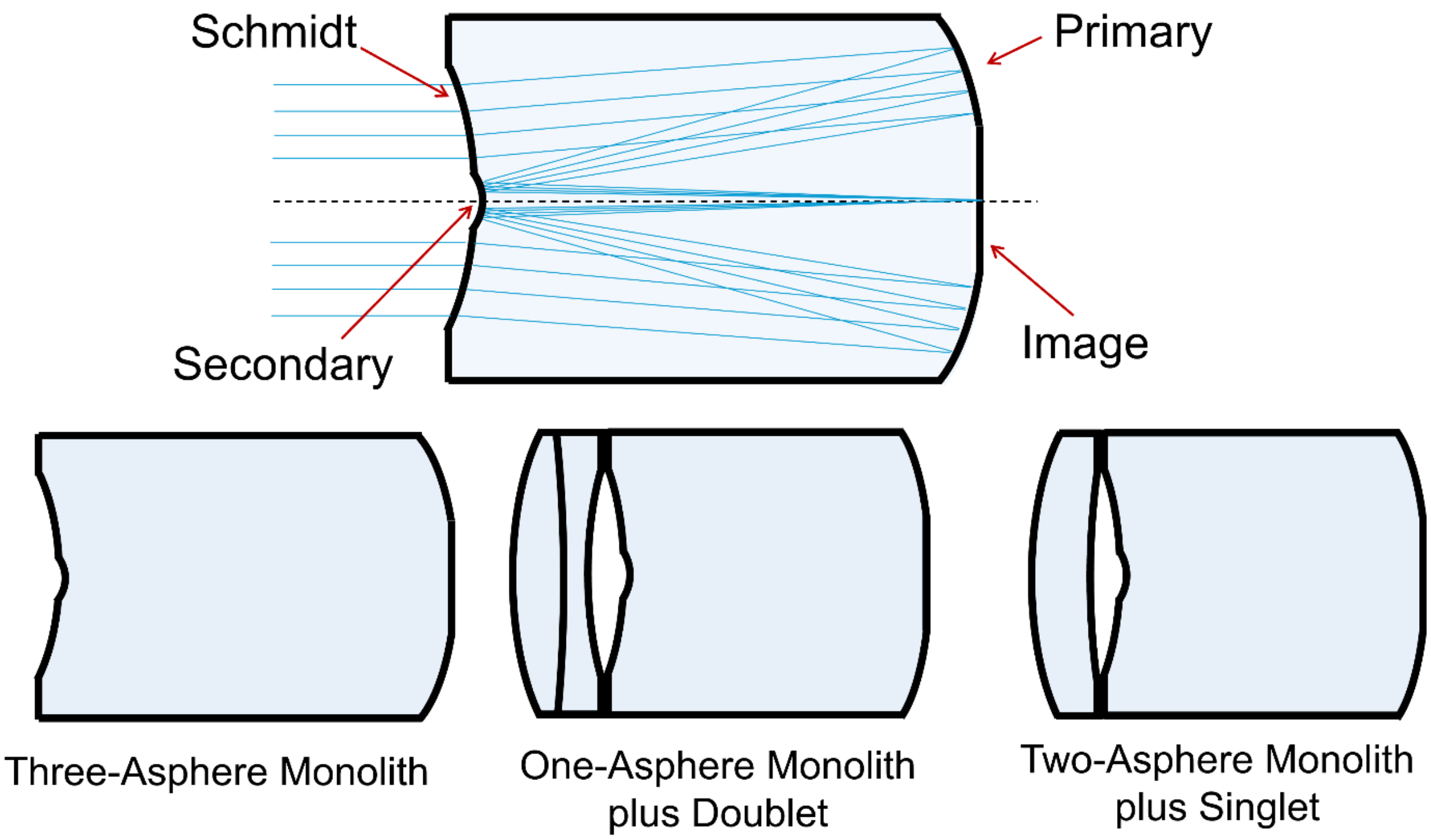

**Figure 2**. Convention for naming surfaces on the monolithic catadioptric lens. The entrance surface is called the "Schmidt" surface, as it is analogous to the Schmidt corrector surface in a Schmidt-Cassegrain telescope. Next, light reflects off the primary mirror, the secondary mirror, and reaches a focus at the image surface. The image surface is flat, so that a slit can be made with photolithography. Three design variants are considered: a three-asphere monolith, a single-asphere monolith with a doublet preceding it, and a two-asphere monolith with an aspheric singlet preceding it.

## 3.2 Three-Asphere Monolithic Lens

The first variant we designed was a single-piece monolith based on a Schmidt-Cassegrain telescope, with three aspheric surfaces and the image plane coincident with the flat back surface of the monolith. Since the exit surface from the monolith cannot be made curved and hence contribute to aberration correction as with other similar monolithic lenses, we needed three aspheric surfaces to obtain well-corrected imaging over the field of view.

Most of the optical work is done by the primary and secondary mirror surfaces, with the weak Schmidt surface providing the necessary correction for off-axis fields. The Schmidt surface has little optical power; otherwise, it would result in larger chromatic aberration. When the weakly-curved Schmidt surface was made aspheric, the spot size was diffraction limited for fields up to 3° off-axis. On the other hand, making the Schmidt surface flat limited the diffraction-limited field of view to 1° off-axis, with coma and astigmatism dominating. Table 3 below compares the spot size for these two design permutations. Meanwhile, we found that the best placement for the aperture stop was at the primary mirror surface.

**Table 3**. Polychromatic RMS geometric spot size demonstrates sizable performance improvement by the Schmidt surface.

| Field Angle | Aspheric Schmidt Surface | Flat Schmidt Surface |
|---|---|---|
| 0° (on axis) | 0.9μm | 3.7μm |
| ±1° | 1.0μm | 2.9μm |
| ±2° | 1.4μm | 7.3μm |
| ±3° | 2.7μm | 17μm |

Figure 3 below shows the optimized lens form ray-trace diagram and the resultant geometric spot size. Performance at λ=0.4μm was slightly improved at the expense of SWIR wavelengths while keeping them diffraction limited. The tangential component of the modulation transfer function (MTF) is considered since the objective lens is coupled to a one-dimensional slit. The MTF has very good performance up to the IMX990 sensor's Nyquist limit of 100 cycles/mm for visible-range wavelengths. Since the Airy disk is ≈3× larger for SWIR wavelengths, their MTF decreases faster and is plotted up to 33 cycles/mm corresponding to the spatial resolution defined by the 15μm slit. Remarkably, this performance holds with hardly any change over -40° to 70°C.

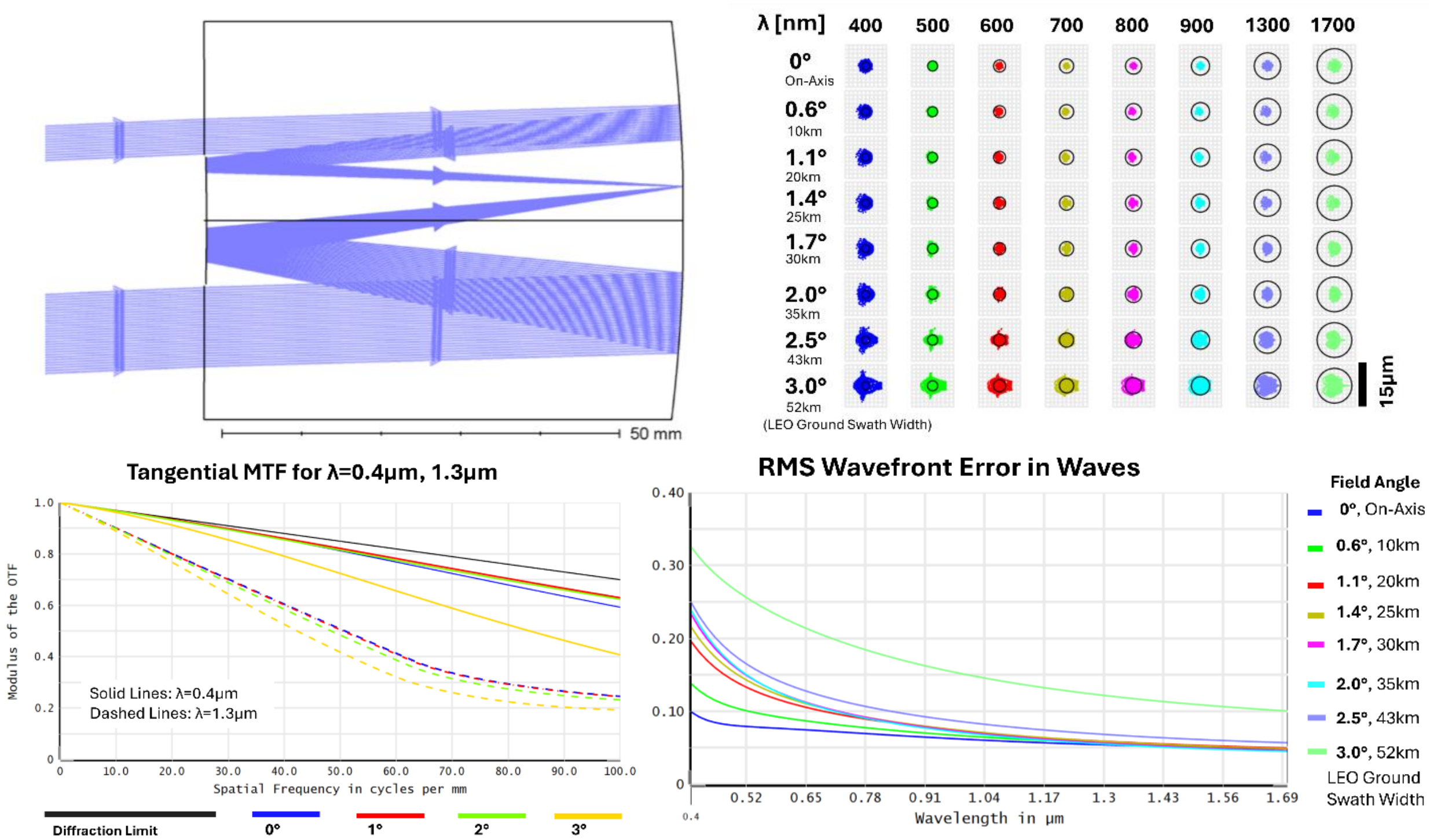

**Figure 3**. Performance of three-asphere monolith. Ray-trace diagram (top left), geometric spot diagram (top right), tangential modulation transfer function (bottom left), and RMS wavefront error (bottom right) at nominal design.

### 3.3 Single-Asphere Monolith and Doublet

A second variant was designed as a "three-piece" lens comprised of the Schmidt-Cassegrain-like monolithic lens and an achromatic doublet placed before its entrance. Two considerations motivated this variant. First, the RMS wavefront error of the all-monolith designs rises sharply toward the short-wavelength end of the band, where the dispersion of fused silica increases rapidly as the ultraviolet absorption edge is approached. A doublet placed in the collimated space ahead of the monolith supplies the opposing dispersion required to flatten this residual and recover performance at λ=0.4µm. Second, the doublet provides additional degrees of freedom for aberration correction, which reduces the number of aspheric surfaces that must be figured on the monolith itself. Since aspheric surfaces on a monolithic blank are both difficult to fabricate and difficult to certify, and are the principal cost driver for this lens class, reducing their number is beneficial for low-cost nanosatellite missions.

We designed the combined doublet-monolith lens system so the doublet can be cemented to the monolith at a flat edge surface surrounding the clear aperture, so overall mechanical rigidity is maintained. Near-diffraction-limited imaging was obtained over the visible, near-infrared, and short-wave infrared regions, with the diffraction ensquared energy within a 15µm × 15µm region on the image plane exceeding 85% in the VIS and NIR range. With the added thickness of the doublet, this design achieves a telephoto ratio of ≈0.7.

When comparing several permutations of this design variant with a single aspheric surface on the monolith, we found that the best location for an aspheric surface is on the Schmidt surface, as opposed to the primary or secondary mirror surfaces. The Schmidt asphere had the highest departure from a best-fit-sphere of 21µm sag, which would require a null-test Fizeau interferometry approach for verifying surface form. This is expected because refracting surfaces contribute much less to wavefront correction compared to reflecting surfaces. By placing the asphere on the secondary mirror surface instead, we reduced the asphere departure to 0.4µm sag while only marginally increasing the spot size. However, the secondary mirror has a tighter absolute tolerance on mid-spatial-frequency error and RMS surface form accuracy as compared to the Schmidt. As before, the aperture stop is best placed at the primary mirror.

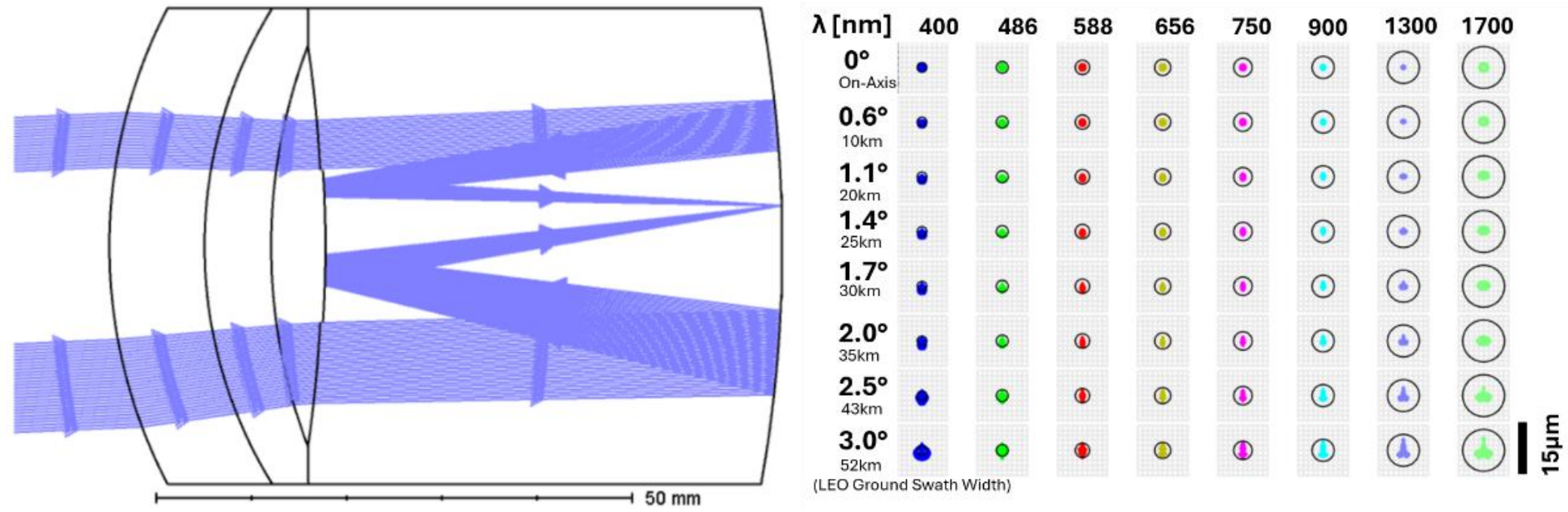

**Figure 4**. Ray trace diagram of the doublet-monolith design variant, and the nominal geometric spot diagram at 20°C. Field of view shown is the half-angle, with ±3° corresponding to 52km of ground swath width at low-Earth orbit altitude of 500km. Airy disk overlaid in each grid, measuring 15µm ×15µm, for scale comparison.

The choice of doublet materials proved to be the most critical one for this design variant. Because the fused silica monolith is nearly achromatic on its own, the doublet is required to supply only a small residual dispersion, which suggests pairing two low-dispersion crown glasses rather than constructing a conventional crown-flint achromat. We initially chose a $CaF_2$ – fused silica pairing, because the thermo-optic coefficients (dn/dT) of the two materials are opposite in sign and comparable in magnitude, at −11.2 and +8.4 $\times10^{-6}$/K respectively at λ=546nm. Modelling the temperature dependence of the refractive index alone, this pairing showed almost no change in RMS spot size between −40°C and 70°C. However, $CaF_2$ has a coefficient of thermal expansion of $\alpha = 18.4\times10^{-6}$/K against $0.5\times10^{-6}$/K for fused silica, and when expansion was modelled together with the index change, the $CaF_2$ – fused silica variant degraded substantially across the operating temperature range and its apparent athermal behaviour disappeared entirely.

It is not straightforward to select glass materials which simultaneously have low dispersion, low CTE, and a specific dn/dT behavior to complement the fused silica monolith. Among oxide crowns, high Abbe number is obtained through fluorine and phosphate content, yet these constituents raise the coefficient of thermal expansion and drive dn/dT negative. A balanced trade-off is made to improve imaging performance at the shorter visible-range wavelengths without subjecting the lens to thermal defocus. We found that FK5HTi provides a good compromise in this regard. It is difficult to find other optical glasses with similarly exceptional

Figure 4 and 5 show the nominal performance of the reoptimized FK5HTi – Fused Silica design. The RMS wavefront error remains below 0.15 waves across the full 0.4-1.7μm band and over the full ±3° field, and the residual now peaks near λ=0.65μm rather than rising steeply toward λ=0.4μm. The short-wavelength spike characteristic of the all-monolith variants has therefore been removed, which was the primary optical motivation for introducing the doublet. Figure 5 shows that this performance is preserved across the operating temperature range: the worst-field peak RMS wavefront error changes only from 0.147 waves at −40°C to 0.139 waves at 70°C, and the shape of the curve is otherwise indistinguishable between the three cases. The doublet design variant also produced the least chromatic focal shift among all three variants studied.

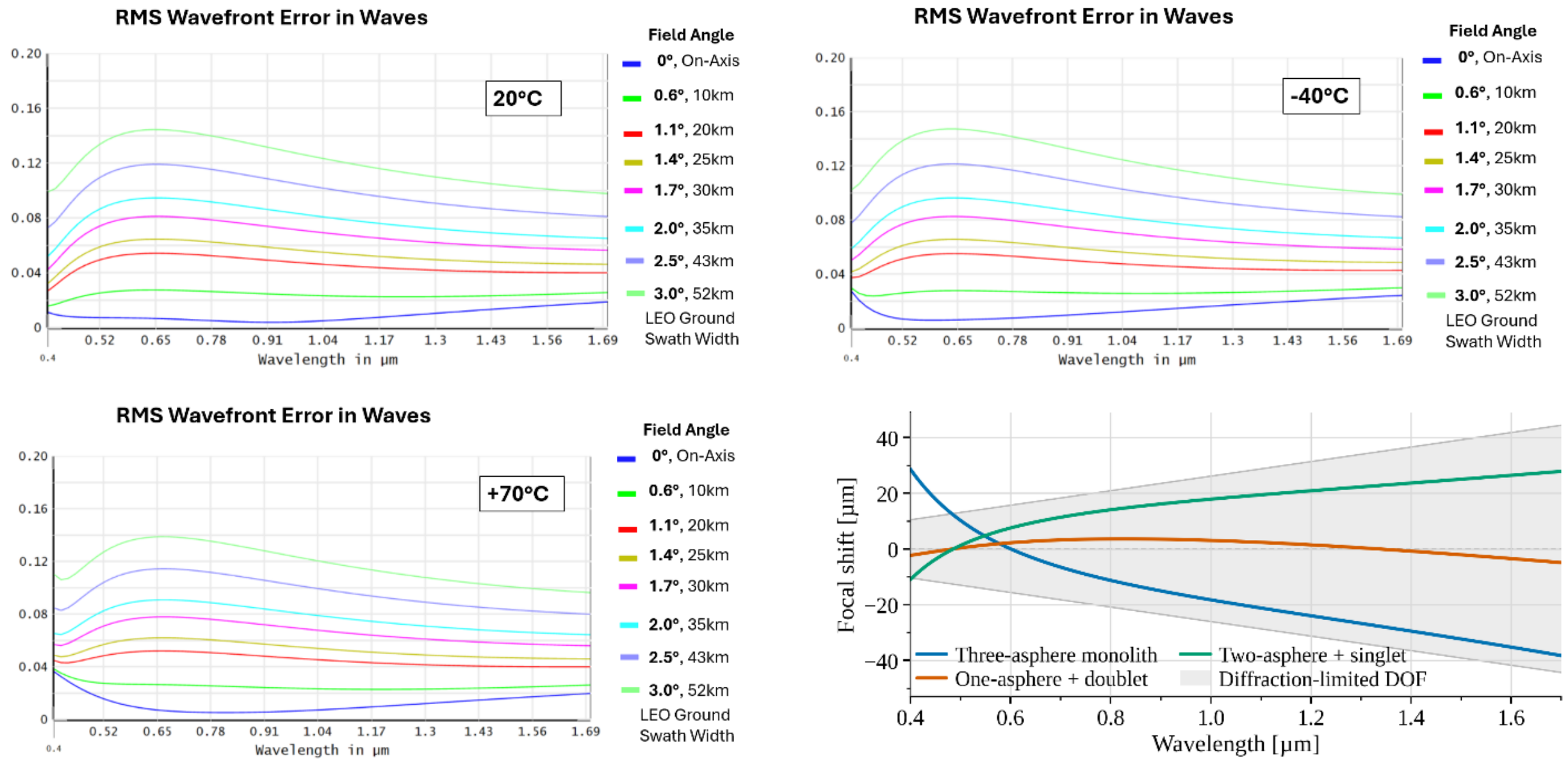

**Figure 5**. RMS wavefront error against wavelength for the FK5HTi – Fused Silica doublet and one-asphere-monolith lens design at −40°C, 20°C and 70°C, for field angles from 0° to 3°. Performance is essentially unchanged across the operating temperature range. Chromatic focal shift across the 0.4–1.7 μm band for the three design variants, against the diffraction-limited depth of focus (shaded), ±2nλN² for F/3 with the image formed inside fused silica.

**Table 4**. Optical and thermal properties of the candidate doublet glasses, ordered by Abbe number. $\Delta n_{abs}/\Delta T$ is the change in refractive index relative to vacuum, averaged over −40/+80°C; the columns "1060.0", "e" and "g" give values at 1060nm, 546nm and 436nm respectively. $\Delta n_{rel}/\Delta T$ is referred to air at 1013 mbar. The thermal power $\gamma = (dn_{rel}/dT)/(n-1) - \alpha$ is the fractional change in element power per kelvin. Fused silica and the selected crown, FK5HTi, are shaded. Glass material property data taken from the Schott glass catalog.

| Glass | $n_d$ | $\nu_d$ | $\alpha_{-30/+70°C}$ | $\Delta n_{abs}/\Delta T$ [$10^{-6}$/K] | | | $\Delta n_{rel}/\Delta T$ at e [$10^{-6}$/K] | Thermal power γ [$10^{-6}$/K] |
|---|---|---|---|---|---|---|---|---|
| | | | | **1060.0** | **e** | **g** | | |
| $CaF_2$ | 1.43385 | 95.23 | 18.4 | −11.4 | −11.2 | −11.0 | −9.9 | −41 |
| N-FK58 | 1.45600 | 90.90 | 13.7 | −7.5 | −7.3 | −7.1 | −6.0 | −27 |
| **FK5HTi** | 1.48748 | 70.47 | 9.2 | −2.9 | −2.5 | −2.2 | −1.1 | −11 |
| **Fused silica** | 1.45844 | 67.83 | 0.5 | +7.9 | +8.4 | +8.8 | +9.8 | +21 |
| N-BK7 | 1.51680 | 64.17 | 7.1 | +0.9 | +1.4 | +1.9 | +2.8 | −2 |
| LF5HTi | 1.58144 | 40.89 | 9.1 | −0.8 | +0.5 | +1.8 | +2.0 | −6 |
| N-SF2 | 1.64769 | 33.82 | 6.7 | +2.2 | +3.7 | +5.5 | +5.2 | +1 |
| N-SF8 | 1.68894 | 31.31 | 8.6 | −0.7 | +0.9 | +3.0 | +2.5 | −5 |

**Table 5**. Comparison of RMS geometric spot sizes weighted equally for 0°, 1°, 2° and 3° in three wavelength ranges, for three permutations of the doublet-monolith design variant. These values consider the impact of dn/dT and CTE combined.

| RMS spot size, averaged [µm] | | | −40°C | | | 20°C | | | 70°C | | |
|---|---|---|---|---|---|---|---|---|---|---|---|
| | | | VIS | NIR | SWIR | VIS | NIR | SWIR | VIS | NIR | SWIR |
| 1 | $CaF_2$ | Fused Silica | 4.66 | 4.80 | 6.88 | 1.68 | 1.82 | 8.03 | 1.52 | 1.53 | 9.92 |
| 2 | N-BK7 | Fused Silica | 1.95 | 2.03 | 3.67 | 1.22 | 1.41 | 3.84 | 0.92 | 0.98 | 4.26 |
| 3 | FK5HTi | Fused Silica | 0.61 | 0.72 | 1.20 | 0.65 | 0.75 | 1.06 | 0.64 | 0.71 | 1.05 |

## 3.4 Two-Asphere Monolith and Cemented Asphere Singlet

The third variant we designed is a “two-piece” lens comprised of the Schmidt-Cassegrain-like monolith with a fused silica aspheric singlet bonded to a flat flange on its front side. Fabrication complexity and cost of the monolithic lens is a significant barrier to its wider adoption for nanosatellite payloads. Our goal with this design was to demonstrate the performance of the monolithic lens can be recovered through robust compensators, thereby relaxing manufacturing tolerances and eliminating the need for expensive metrology solutions. We introduce the design and compensator definitions presently and discuss results of our tolerance analysis in Section 3.5.

Other studies of tolerances in monolithic catadioptric lenses have shown the most critical tolerances are radial decenters between annular zones and the axial separations between surfaces [40]. In a monolith, an axial error changes the separation between powered surfaces and defocuses the image directly. Simultaneously, meeting a stringent center-thickness tolerance is difficult because thickness is not controlled directly during lens surface grinding and polishing: material is removed to achieve the target radius and surface form, and the final thickness is whatever remains once those requirements are met. Deterministic polishing methods such as magnetorheological finishing hold center thickness more tightly but are prohibitively slow over the apertures and removal depths typical of this lens class. While a common approach to mitigate defocus when the image plane lies in air is to adjust the axial position of the imaging sensor, this compensator is susceptible to mechanical misalignment and thermo-elastic drift. Monolithic lenses are also sensitive to lateral errors between annular surface regions, which manifest as coma aberration. Displacing a rotationally symmetric surface laterally by $\delta$ changes the sag to $z(r) - \delta \cdot \partial z/\partial x$. For an asphere with a fourth-order term $a_4 r^4$, the induced shape is $4a_4 r^3 \delta \cos\theta$, which is well-recognized as coma aberration that grows linearly with $\delta$ and with the strength of the asphere. Re-chucking the monolith results in large tilt and decenter tolerance between powered aspheric surfaces on its front and back sides, and those tolerances reduce the well-corrected field of view.

The lens variant we designed addresses these sensitivities with the following two compensators. Radial decenter between aspheric surfaces can be corrected for by carefully decentering the aspheric singlet relative to the monolith, until the coma aberration approximately cancels out. The flat flange on the monolith is designed to be outside its clear aperture, so it can be used as a mating surface with optical precision for flatness and parallelism. Meanwhile, center thickness tolerance can be compensated either by trimming the flat-polished back side of the monolith, or by shimming it with a flat fused silica disk of a carefully controlled thickness. Therefore, we intentionally design the monolithic structure to have a boss, or a protrusion, over its exit aperture. The boss can be machined into the monolith by removing material around it, or by hydroxide-catalysis bonding a fused-silica flat onto the monolith.

The nominal design’s performance is shown in Figure 6 below. Diffraction-limited imaging quality is achieved over the VSWIR range, and the RMS spot size remains virtually unchanged over -40°C to 70°C due to the inherently athermal behavior of fused silica. By allowing one surface of the singlet to be aspheric, we can reduce the number of aspheric surfaces on the monolith from three to two while retaining excellent correction for off-axis fields. Previous designs have shown the Schmidt surface typically has large aspheric sag departure which makes the metrology challenging, so we opted to keep aspheric prescription on the primary and secondary surfaces. We placed constraints on the maximum slope departure and peak-to-valley sag departure of the aspheres relative to their best-fit-spheres, computed as the minimum-RMS with offset as relevant for surface form metrology. The aspheres did not exceed 1.1µm/mm slope departure and 3µm sag departure over their clear apertures. Moreover, this design did not leverage even polynomial asphere terms, relying only on conic constants for correction. The simpler overall prescription enables more accurate metrology and hence fabrication of these surfaces.

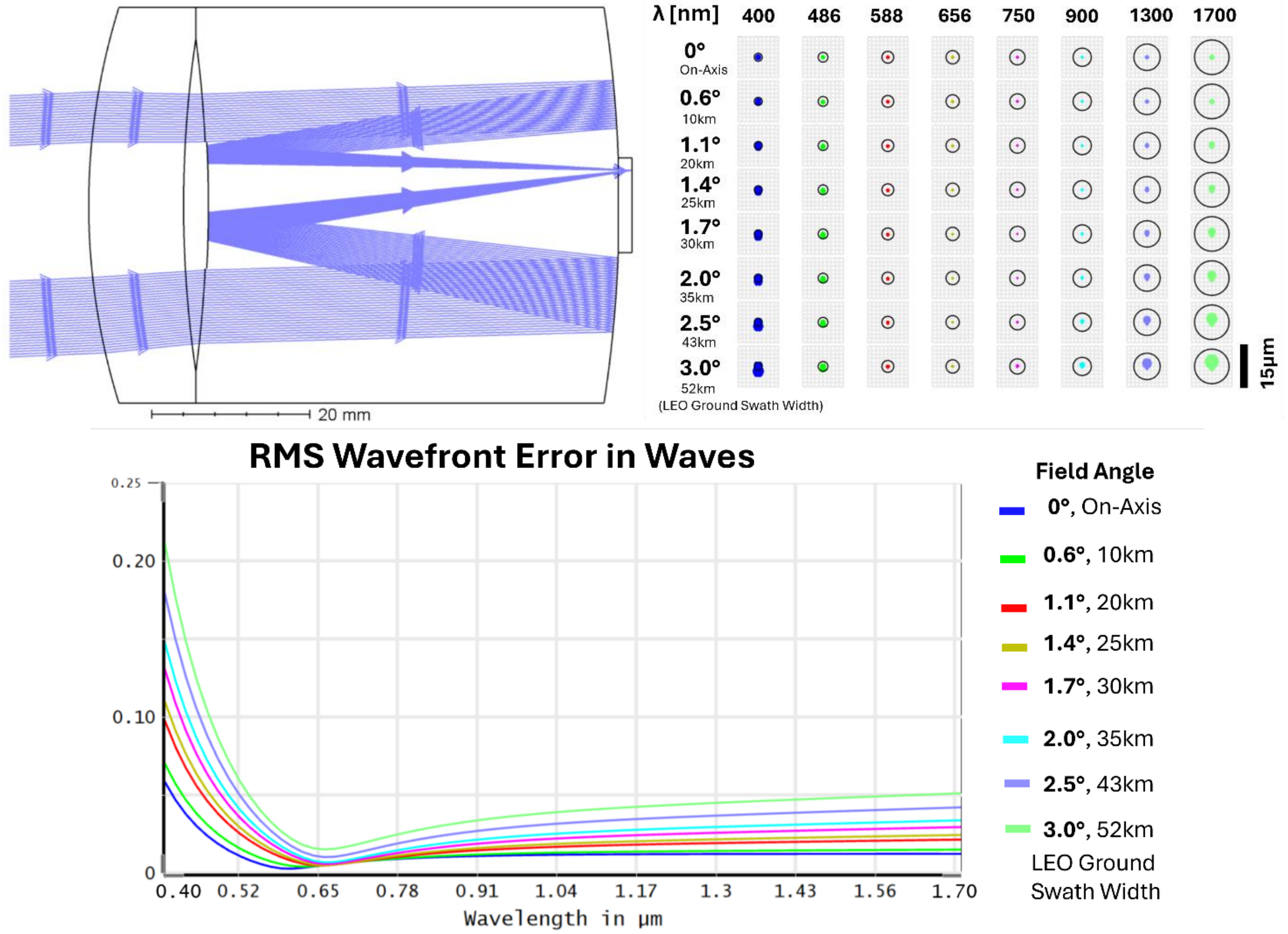

**Figure 6**. Performance of aspheric Fused Silica singlet and two-asphere-monolith lens design. Ray-trace diagram (top left), geometric spot diagram (top right), RMS wavefront error (bottom center), all at nominal performance.

The compensators proposed for this design variant are robust and de-risk the manufacturing of the overall monolithic lens. The aspheric singlet preceding the monolith is a standard optical lens that can be readily fabricated and inspected using a standard Fizeau interferometer with transmission sphere without nulling optics. Meanwhile, center thickness tolerance can be much better controlled when the polished surface is flat as opposed to curved, so the flat boss surface can serve as an adequate focus compensator. The boss protrusion would be intentionally oversized and polished down to the right thickness. Alternatively, a fused silica flat can be separately polished to the right thickness and hydroxide-catalysis bonded onto the monolith.

An initial estimate of the compensator adjustment range can be made before proceeding with a rigorous Monte Carlo simulation-based analysis. If the asphere on the singlet has a similar power to an asphere on the monolith, then any fabrication-engendered decenter on the monolith's asphere can be compensated with a 1:1 travel of the singlet in the opposite direction. Meanwhile, the F/3 lens has a numerical aperture of 0.16 in air but 0.11 inside fused silica, which means the beam's depth of focus $z_R = \frac{\lambda_0}{\pi \cdot \mathrm{NA}^2}$ is almost double of that in air. This has the advantage of reducing the sensitivity of center thickness trims at the expense of a larger compensator range. If the beam's $1/e^2$ intensity diameter was 15μm, the Rayleigh range in fused silica is almost 200μm and we can expect the compensator range to be similar in magnitude.

## 3.5 Tolerance Analysis for Two-Asphere Monolith and Cemented Asphere Singlet

A tolerance analysis is conducted on the design variant in which a fused silica aspheric singlet lens is mated to the fused-silica monolith. Table 6 below summarizes the values used for the tolerancing model, which treats the optical axis of the Schmidt surface as the datum axis. Since the Schmidt and secondary mirror surfaces can be machined in one chucking, they enjoy relatively tight tolerance and inter-surface registration afforded by the CNC equipment. On the other hand, tolerances on the tilt and decenter of the primary mirror and the center thickness between the primary and Schmidt surfaces are larger because the optic must be re-chucked. Form accuracy tolerance is tighter on the secondary mirror because it has weak asphere departure and is more susceptible to surface form irregularity.

**Table 6.** Summary of tolerance quantities and Monte Carlo simulation performance and compensator statistics.

| Schmidt Surface | | Statistics | |
|---|---|---|---|
| Radius | ±0.05% | Performance | Average RMS Spot [µm] |
| Surface Irregularity | λ/20 RMS | Nominal | 0.701 |
| Thickness to Primary | ±50µm | Mean | 0.822 |
| Thickness to Secondary | ±10µm | Standard Deviation | 0.067 |
| Primary Mirror Surface | | Worst Case | 1.120 |
| Radius | ±0.05% | Compensator Range | |
| Surface Irregularity | λ/20 RMS | Thickness Trim | Thickness [mm] |
| Tilt | ±0.02° | Mean (Nominal) | 0.998 (1.000) |
| Decenter | ±20µm | Standard Deviation | 0.073 |
| Secondary Mirror Surface | | Min-Max Range | 0.78 to 1.20 |
| Radius | ±0.05% | Singlet Decenter | Displacement [mm] |
| Surface Irregularity | λ/40 RMS | Mean (Nominal) | 0.001 (0.000) |
| Tilt | ±0.01° | Standard Deviation | 0.038 |
| Decenter | ±10µm | Min-Max Range | -0.11 to 0.12 |

With these tolerances and compensators, we can keep the RMS geometric spot size averaged over wavelengths 0.4µm-1.7µm fields 0° - 3° to within 1.1µm, whereas the nominal RMS geometric spot size is 0.7µm with the same averaging. The spot size is maintained over -40°C to 70°C. Since the aspheric singlet is a standard optical lens form, we do not consider contributions from its potential manufacturing tolerances, although they can be readily absorbed by the compensators. Figure 7 below shows the performance of the toleranced lens with compensators applied.

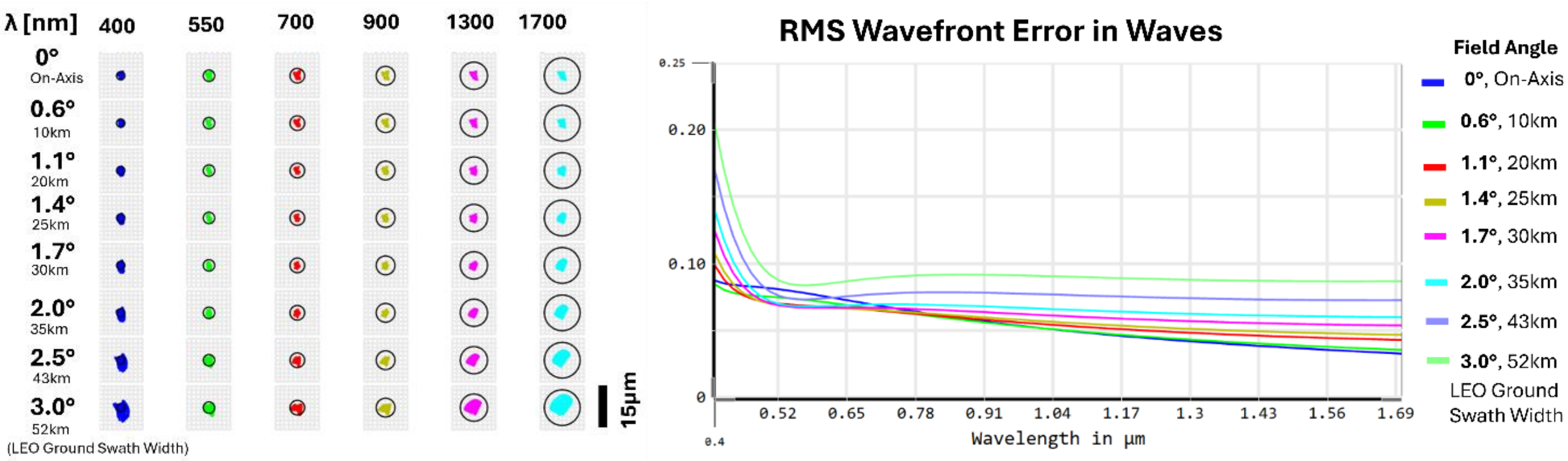

**Figure 7**. Worst-case Monte Carlo performance of the aspheric Fused Silica singlet and two-asphere-monolith lens design, after applying compensators.

The back plane thickness trim contributes the most to performance recovery for this tolerance model. Varying the thickness of the boss compensator within ±10µm results in minimal change to the RMS spot size performance, even though ±1µm can be readily accomplished. The compensator range of 420µm agrees with our earlier estimates based on the beam's Rayleigh range.

Without the compensators on back plane thickness trim, the largest contributors to performance degradation were the center thickness between the Schmidt and primary mirror surfaces, together with the radius of curvature of the primary mirror. To a lesser extent the system was affected by the radius of curvature of the Schmidt and secondary mirror surfaces, their center thickness tolerance, and their surface irregularity. With the back plane thickness trim compensator enabled, performance degradation came from the tilt and decenter on the primary and secondary mirror. In fact, doubling the tilt and radial decenter tolerances does not result in a large penalty to performance, with the worst-case RMS geometric spot size increasing to 1.5μm. On the other hand, doubling the tolerance on all center thicknesses and radii of curvature results in a larger penalty to performance, with the worst-case RMS geometric spot size increasing to 1.7μm and the thickness trim compensator range roughly doubling to 900μm. Without the back plane thickness trim compensator, halving the tolerances on radius of curvature and center thickness does not come close to restoring nominal performance, with Monte Carlo simulations predicting worst-case RMS geometric spot size of 8.0μm and mean spot size of 2.5μm.

The compensation methods are clearly powerful for this optical design yet can also be readily implemented. The back plane thickness can be trimmed in the following way. When a microscope objective is aligned to be in focus at the current back surface and a collimated plane wave beam is launched into the lens, the point spread function (PSF) of the lens can be measured. The initial position of the singlet can be located through coarse alignment, and the back plane surface polished down incrementally until the sharpest PSF is achieved. By measuring the double-pass transmitted wavefront through the whole lens (singlet plus monolith) with a standard Fizeau interferometer fitted with a transmission sphere and a reference flat, the position of the singlet can be precisely adjusted to minimize wavefront error.

## 4. CONCLUSION

Monolithic catadioptric lenses fold a long focal length into a short physical envelope and fix the separations between their powered surfaces at fabrication, so the alignment cannot drift on orbit or be disturbed by launch. Fused silica adds a low expansion coefficient, broad VSWIR transmission, and resistance to radiation darkening, making this lens form well matched to nanosatellite payloads that are constrained simultaneously in volume, mass, and ruggedness. The same architecture is demanding to manufacture. The blank must be re-chucked to figure surfaces on both faces, which introduces tilt and decenter between powered aspheres; center thickness is not controlled directly during grinding and polishing, but is whatever remains once radius and surface form are met; and a figure error on a glass-immersed mirror contributes roughly six times the wavefront error of the same error on a refracting face.

We proposed two ideas to make this lens form more manufacturable and robust. The first is an integrated slit that would serve as the entrance to a subsequent spectrograph in a pushbroom hyperspectral imager. Constraining the image plane to coincide with the flat-polished back surface of the monolith allows the slit to be lithographically patterned onto that surface, which removes the slit-to-objective alignment difficulties characteristic of mechanically mounted air slits. The second is a pair of compensators that absorb the dominant fabrication errors without breaking the monolithic architecture: a lateral slide of the cemented aspheric singlet, which cancels the coma introduced by re-chucking, and a trim of a raised boss around the exit aperture, which sets focus against center thickness error on a flat and readily measured surface.

The design combining these ideas achieves f=100mm at F/3 over 0.4μm–1.7μm and ±3° of field, with diffraction-limited imaging averaged over wavelength and field. Both aspheres on the monolith are conic-only, departing less than 3μm in sag and 1.1μm/mm in slope from their best-fit spheres, so they can be certified on a standard Fizeau interferometer without nulling optics. After fabrication tolerances are applied, near-diffraction-limited performance can be restored with a thickness trim range of 420μm and a singlet decenter within ±0.12mm, and this performance is essentially unchanged from -40°C to 70°C.

Several avenues for further work remain open. The singlet was optimized for nominal imaging performance rather than for compensation authority, and a design that trades nominal wavefront error against compensator range should widen the tolerance envelope further. The tolerance model should also be extended to include the singlet's own fabrication errors, which we assumed the compensators would absorb. Stray light analysis has not been performed, yet it is especially important for this lens class. Because the imaging and stray-light paths share the same block, baffles must be drilled into the blank, and since a baffle bore vignettes off-axis fields, baffle geometry and usable field of view must be optimized jointly. Finally, a more rigorous analysis considering the impact of thermal gradients expected from orbital solar loading on the higher-order wavefront error can be performed.

## ACKNOWLEDGEMENTS

The authors thank the University of Toronto Engineering Society and the University of Toronto Students' Union for their support, and Ansys for their sponsorship of optical design software.